# Combining optical diffraction tomography with imaging flow cytometry for characterizing morphology, hemoglobin content, and membrane deformability of live red blood cells

Yu-Hsiang Chang,[1] Yang-Hsien Lin,[1] Kung-Bin Sung[1,2,3,*]

[1]Graduate Institute of Biomedical Electronics and Bioinformatics, National Taiwan University
[2]Department of Electrical Engineering, National Taiwan University
[3]Molecular Imaging Center, National Taiwan University

*Kung-Bin Sung, E-mail: kbsung@ntu.edu.tw

**Abstract:** Integrating optical diffraction tomography with imaging flow cytometry enables label-free quantifications of the three-dimensional (3D) morphology and hemoglobin content of red blood cells (RBCs) in their natural form. Self-rotation of RBCs flowing in a microfluidic channel has been utilized to achieve various projection directions for 3D reconstruction. However, the practicality of this technique has not been sufficiently studied. We improved the accuracy of estimating the rotation angle of RBCs and demonstrated 3D reconstructions of both healthy and glutaraldehyde-treated RBCs. Results showed the capability to quantify changes in RBC morphology, hemoglobin content, and membrane fluctuations generated by glutaraldehyde treatments, demonstrating the potential to detect changes frequently present in various RBC membrane disorders.

## 1 Introduction

Healthy red blood cells (RBCs) are in the shape of biconcave discs. This three-dimensional (3D) morphology is crucial for the primary function of RBCs, which is the transportation of oxygen. Deviations in the morphology from the optimal discocytes often hinder normal functions of RBCs and, therefore, are targets for diagnostic methods such as flow cytometry and blood smear [1]. In flow cytometry, which has high throughput but requires bulky instruments, the size of individual RBCs can be estimated from impedance or forward light scattering measurements. However, detailed 3D morphology information is not available. On the other hand, a blood smear provides two-dimensional (2D) images of stained RBCs, revealing cellular morphology and intracellular distributions of hemoglobin. However, both the staining and image interpretation are more demanding regarding resources. Therefore, it is usually performed to facilitate diagnosis rather than for screening.

Quantitative phase imaging (QPI) is an emerging technique that quantifies the intrinsic phase contrast in living cells [2-4]. Moreover, 3D reconstruction of the internal refractive-index (RI) distributions of living cells can be achieved using multiple 2D QPI of the cells at different illumination angles through beam scanning [5, 6], sample rotation [7], or sample flowing [8]. 3D morphology of RBCs from patients with various blood-related diseases has been quantified and has the potential to facilitate the screening and management of the conditions [9-11]. Since QPI does not require labeling and can be implemented with relatively simple and cost-effective hardware, it is suitable for making widespread point-of-care or screening devices. However, tomographic imaging has been achieved mainly by scanning the incident beam on stationary RBCs with low throughput.

A common strategy to increase the data acquisition rate is integrating a microfluidic channel into the QPI instrument and acquiring images of cells or organisms flowing continuously in the medium [12-14]. To acquire QPI data for tomographic reconstruction, Merola *et al.* proposed carefully controlling the flow of RBCs through a microfluidic channel to make them rotate continuously. The rotation angles needed for reconstruction are determined by fitting the RBCs' projected phase images to Zernike polynomials [15]. This method has higher throughput than other works where projection directions are varied by scanning the illumination beam [16], or rotating RBCs with optical traps [17] or dielectrophoresis [18] in microfluidic channels. Yet another method continuously records forwardly scattered light from a line region illuminated by a convergent beam while cells flow across the line [19, 20]. The angular range of projections is limited by the collecting cone of the objective lens and cannot reach a full 180° rotation as in the self-rotation method proposed in [15], resulting in lower spatial resolution.

Although 3D RI tomograms have been obtained from continuously flowing RBCs, the method's reliability has not been demonstrated to show its practicality. Specifically, only four abnormally shaped RBCs were reported in [15], and one RBC tomogram was shown in [21]. Therefore, we evaluated the performances of estimating the rotation angle of RBCs and the subsequent 3D reconstruction using simulated RBC QPI projections. With improved accuracy in the rotation angle determined by a modified procedure, we applied optical diffraction tomography [22, 23] to reconstruct 3D RI tomograms of live RBCs flowing in a custom-made microfluidic device. The rotation of RBCs was confined to the central region of the sample channel by a sheath flow, and complex field images of the RBCs were continuously acquired by off-axis digital holographic microscopy (DHM). To test the capability of the developed tomographic imaging flow cytometer in quantifying changes in RBC 3D morphology and hemoglobin content, we treated healthy RBCs with glutaraldehyde that stiffened the membrane and reduced the deformability of RBCs. The treatment was intended to imitate trends observed in abnormalities such as hereditary spherocytosis, hereditary elliptocytosis, and metabolic syndrome [24].

## 2   Materials and Methods

*2.1 DHM setup*

Quantitative phase images of RBCs flowing in a microfluidic channel were acquired by a DHM setup whose schematic diagram is shown in Fig. 1. Off-axis interferometry is implemented in a common-path arrangement in which a uniform reference beam is generated by a transmission grating [25]. A 532-nm laser is spatially filtered (pinhole diameter 10 μm), expanded, and focused to the back focal plane of a condenser lens to provide nearly plane-wave illumination at normal incidence on the channel. A water-immersion objective lens (LUMFLN 60X, NA 1.1, Olympus) collects light scattered by the RBCs. It forms an intermediate image that is relayed by a 4f lens system onto a complementary metal-oxide semiconductor camera (VC-12MX-M 180, Vieworks Co., Ltd.) with a transverse magnification of about 84. The field of view is about 200 μm × 200 μm. Unscattered illumination beam is also collected by the objective lens, magnified, and relayed onto the camera as a uniform beam. A grating (300 grooves/mm UV Transmission Grating, Dynasil Corporation) splits the transmitted light (both scattered and unscattered by the RBCs) into multiple beams, and only the 0 and -1 orders are allowed to pass a clear aperture located at the Fourier plane of the intermediate image. The grating is moved a few millimeters away from the intermediate image plane along the optical axis to shear RBC images of the -1-order beam relative to those of the 0-order beam. The shearing direction is perpendicular to the sample flow direction, and the shearing distance is adjusted so that RBC images in the 0-order (sample) beam overlap with an empty region without flowing RBCs in the -1-order (reference) beam [25, 26]. This strategy for creating a nearly common-path uniform reference beam has two advantages over spatial filtering with a pinhole, as used in original diffraction phase microscopy [27]. First, it is less susceptible to misalignment and movements of optical components such as the pinhole. Second, the grating can be chosen to have roughly equal efficiency in the -1 and 0 orders to achieve higher fringe contrast [28, 29]. The lateral resolution of the optical imaging system was measured to be 0.26 μm under white light illumination and is close to the theoretical prediction of diffraction-limited systems. Images were acquired at 180 frame/s with an exposure of 1 ms for results reported here.

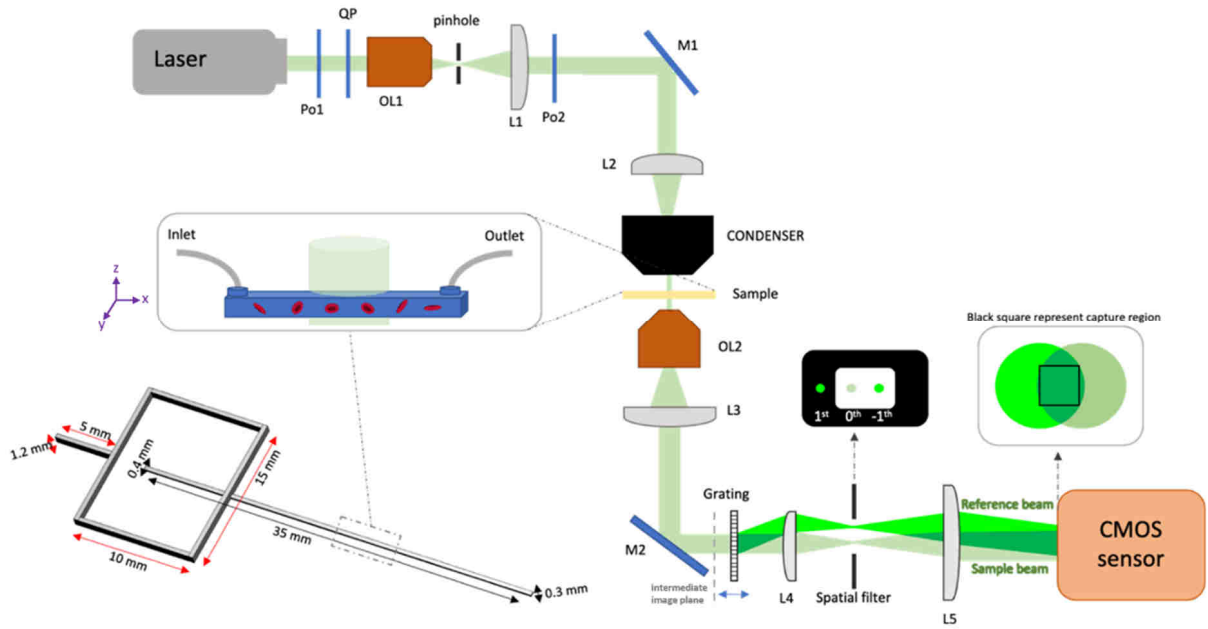

**Fig. 1** Schematic diagram of the imaging cytometer. Po1 & Po2: linear polarizers; QP: quarter waveplate; OL1 & OL2: objective lenses; M1 & M2: mirrors; L1~L5: positive lenses. Dashed black lines adjacent to the grating indicate the location of an intermediate image of the sample.

## 2.2 Design and fabrication of the microfluidic device

To facilitate the rolling of RBCs for at least one revolution within the field of view (FOV) and prevent RBCs from flowing through out-of-focus regions, we adopted a 3D hydrodynamic focusing method to help confine RBCs near the bottom of the channel [30]. A schematic diagram of the microfluidic device is shown in Fig. 2(a). The 0.3 mm thick central sample channel was shallower than the 1.2 mm thick sheath channels, and the flow rate of the sheath stream was faster than that of the sample stream to achieve 3D hydrodynamic focusing effects. The microfluidic channel was fabricated by casting polydimethylsiloxane (PDMS) against a positive micro-milled aluminum mold and boding the cured PDMS slab onto a coverglass with oxygen plasma surface treatment. Syringe pumps were used to push the sample and sheath fluids with a flow rate of 5 μl/min and 15-20 μl/min, respectively. As seen in Fig. 2(b), the faster sheath flow occupied most of the channel height and successfully confined the sample flow to near the bottom of the channel.

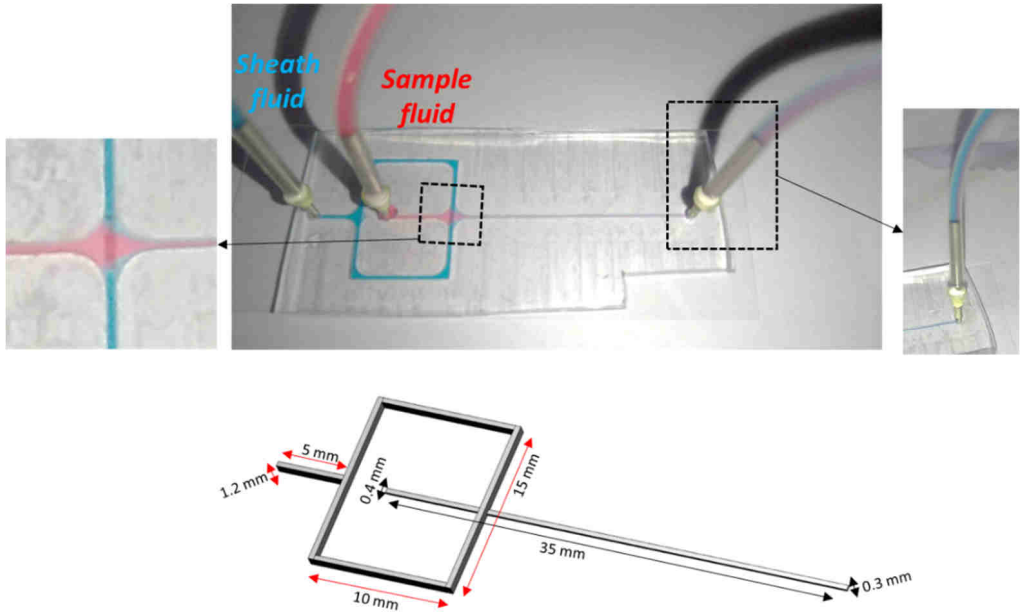

**Fig. 2** (a) Schematic of the microfluidic 3D focusing device. (b) A photograph of the microfluidic device filled with red (sample fluid) and blue (sheath fluid) solutions. The inset on the left shows the horizontal confinement of the sample flow, and the inset on the right shows the vertical confinement of the sample flow.

*2.3 Image Processing, Phase Reconstruction, and Phase Unwrapping*

Since many interference images were acquired during the time when an RBC flowed across the FOV, individual RBCs in the raw images were automatically detected by an image cascade network [31] and cropped to speed up the processing. Training of the network is described as follows. First, a mean intensity image was calculated from the whole series of continuously recorded images and subtracted from each recorded image. Second, local nonuniformity in intensity was alleviated by contrast-limited adaptive histogram equalization. Third, we manually labeled more than 13,000 RBCs in about 2,200 recorded interference images and randomly split the labeled RBC images into a training set and a validation set with a 9:1 ratio to train the network using PyTorch. The implemented network on a personal computer equipped with an Intel Core i5-9400F CPU and NVIDIA GeFore RTX 2060 GPU achieved a processing speed of 18 frames/s for 1024x1024 pixels/frame, and an average intersection over union (IoU) of 0.822.

We reconstructed quantitative phase images of individual RBCs by the spatial-frequency filtering method [32, 33]. The method consists of bandpass filtering the recorded images around the spatial frequency of interference fringes, performing inverse Fourier transformation, taking the argument of the complex image, and subtracting a background phase image of the same FOV but without RBCs. The amplitude image of each RBC was also obtained from the inverse Fourier transformation. The $2\pi$ phase ambiguity issue was solved by a fast 2D phase-unwrapping algorithm [34]. Moreover, during the movement of RBCs in the channel, the location of RBCs along the optical axis is not always constant. To ensure high image quality, we applied numerical refocusing to obtain the best-focused RBC phase image. We propagated the complex-field (i.e., amplitude and phase) image with the angular spectrum method and searched for the focus-shifted phase image with the maximum Tamura coefficient [35]. After refocusing the complex images, we calculated the center of mass of each RBC phase image and aligned the centers of all phase images belonging to each RBC for subsequent 3D reconstruction.

## 2.4 Rotation Angle Determination

The orientation of an RBC in the microfluidic channel can be described by two rotation angles designated as θ and γ hereafter. As illustrated in Fig. 1, the x-axis and z-axis align with the flow direction in the channel and the optical axis of the DHM setup, respectively. θ indicates the angle of rotation around the y-axis when the RBC rolls continuously due to the flow in the channel, and γ refers to the rotation angle around the z-axis due to imbalanced flow speeds at two sides of the RBC. The γ of each RBC phase image was first determined by elliptical fitting of the RBC, and the RBC image was numerically rotated so that the long axis of the RBC was aligned with the y-axis. Since RBCs oriented close to θ = 0°, i.e., with disk-like appearance, have high levels of radial symmetry and are challenging to determine γ, we only performed the elliptical fitting on RBCs with their major axis at least five pixels longer than their short axis. γ values of RBC phase images without the elliptical fitting were determined by interpolation.

After re-rotating RBC images around the z-axis, we determined the rolling angle θ of each RBC image based on Zernike polynomial fitting of RBC phase images [15] with some modifications. The relationship $C_4 \propto \cos2(\theta)$ was used in [15] to calculate $\theta = \cos^{-1}[(C_4)^{0.5}]$, where $C_4$ is the Zernike coefficient of defocus. However, based on our experimental results of RBCs moving in the channel, RBC phase images showed both time-varying intracellular mass distributions and nonideal biconcave discs during the movements, which resulted in RBCs with the same rolling angle corresponding to different $C_4$ values. Therefore, we modified the previous method to improve accuracy in the rolling angle estimation. Specifically, RBC phase images in the orientations around θ=90° and θ=0° were first identified.

RBCs with θ around 90° had their axes of symmetry aligned with the x-axis and were identified by finding frames with a local maximum of $C_{13}$ in the image stack of each RBC, where $C_{13}$ is the Zernike coefficient of horizontal secondary astigmatism. Similarly, RBCs with θ around 0° had their axis of symmetry aligned with the z-axis and were identified by finding frames with a local maximum of $C_4+1/C_{12}$ in the image stack of each RBC, where $C_{12}$ is the Zernike coefficient of primary spherical aberration. The Zernike polynomials used are illustrated in Fig. 3, which show similarity to phase images of RBCs in orientations of θ=0° and θ=90°. After the frames with θ = 90° and θ = 0° were identified, the whole stack of RBC images was divided into segments of θ = 0°-90° and θ = 90°-0°, as illustrated in Fig. 4. Within each segment, we assumed that the frame identified as θ = 90° has the minimum $C_4+C_5$ and that identified as θ = 0° has the maximum $C_4+C_5$. $C_5$ is the Zernike coefficient of horizontal primary astigmatism. Frames with $C_4+C_5$ values below the frame specified as θ = 90° or above the frame identified as θ = 0° within the same segment were deemed outliers and removed from subsequent processing. Finally, the $C_4+C_5$ value of every frame within the segment was normalized to a range between 0 and 1, and the rolling angle was calculated as $\theta = \cos^{-1}[(C_4+C_5)^{0.5}]$.

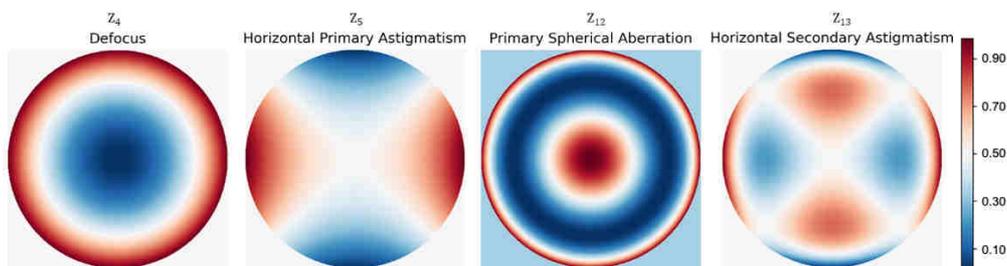

**Fig. 3** Normalized Zernike polynomials that are used to fit RBC phase images for determining the rotation angle θ.

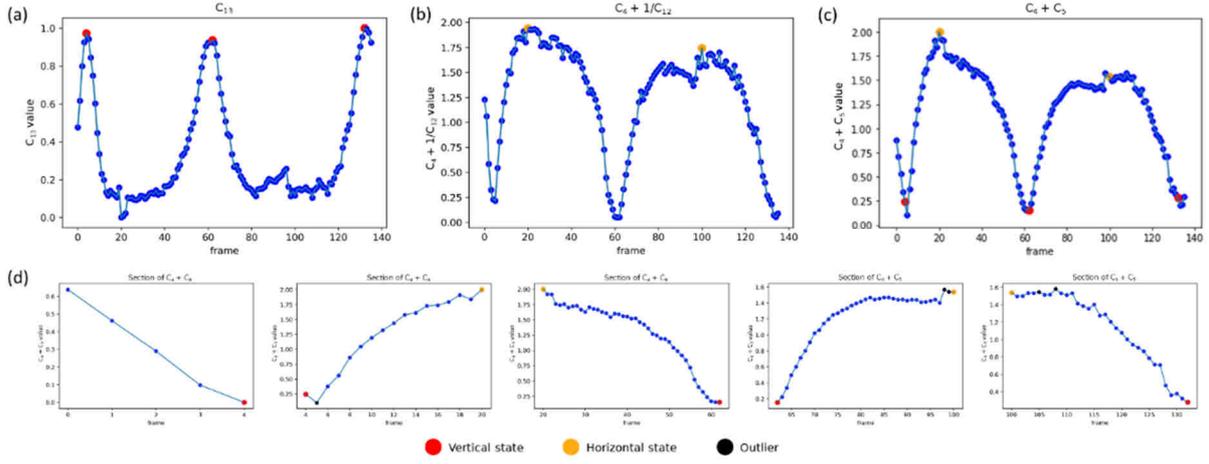

**Fig. 4** An example of estimating rolling angles of a stack of RBC phase images. (a) Frames with a local maximum of $C_{13}$ were identified to be oriented at $\theta = 90°$ (red circles); (b) frames with a local maximum of $C_4+1/C_{12}$ were determined to be oriented at $\theta = 0°$ (yellow circles); (c) $C_4+C_5$ of all frames in the image stack, and (d) segments of frames with $\theta$ between $0°$ and $90°$ as indicated in (c). Black circles are outlier frames that had $C_4+C_5$ above the yellow circles ($\theta = 0°$) or below the red circles ($\theta = 90°$) and were excluded from further processing.

## 2.5 Reconstruction of 3D RI Maps

3D RI distributions of individual RBCs were reconstructed from about 200 complex-field images by backpropagation based on Fourier diffraction tomography under Rytov approximation [36]. Since the RBC field images were only obtained under uniaxial rotation of RBCs about the y-axis, the 2D Fourier transform of acquired field images filled a horn torus-like shape in the spatial frequency domain [23]. Artifacts generated by this missing cone problem were reduced by applying positivity and spatial constraints iteratively. In addition, total variation minimization (TVmin) was used to smooth intracellular RI variations while preserving RBCs' edges in the reconstructed images [37]. The procedure for processing a reconstructed 3D RBC RI image is as follows. We applied the TVmin step to the reconstructed image ten times to approximate the spatial extent of the RBC using Otsu thresholding. Then, we performed morphological opening and closing to create a 3D mask for the RBC. Subsequent TVmin and positivity operations were confined within the mask until the total variation could not be improved, or a maximum iteration was reached.

## 2.6 Evaluation of methods for determining rotation angles

To evaluate the accuracy of the estimated rotation angles by the proposed method, quantitative phase images of RBCs were generated from experimentally acquired and reconstructed 3D RI tomograms [11] as testing data. We took 3D RI tomograms of 15 RBCs and sampled 200 combinations of $\theta$ and $\gamma$ for each RBC to generate test phase images by projection. Increments of $\gamma$ between adjacent frames were randomly sampled from the range of $\pm 5°$. Increments of $\theta$ between adjacent frames were randomly sampled from the range of $\pm 5°$ when $\theta$ was between $-30°$ and $30°$, and from the range of $3°\sim8°$ at other $\theta$ values. This setting was determined based on observations of the relative occurrence rate of RBC movements in the recorded interference images (see visualization 1 for an example). Two types of artifacts were digitally added to the test images to imitate artifacts commonly seen in experimentally measured phase images. The first artifact was temporal fluctuations in the total phase, which were found to be about 9% of the mean value. We randomly sampled from a normal distribution with a 9% standard deviation, calculated the corresponding noise in the total phase, and evenly distributed the noise to all the pixels in an RBC phase image. The second artifact was the warping of RBC phase images, which was achieved by shifting pixel positions row-by-row using

$$x' = S \times \sin\left(\frac{2\pi x}{W}\right) + x, \tag{1}$$

where *x* is the original location of a pixel in the x-axis, *x'* is the new location of the pixel, S is a constant to adjust the amount of warping, and *W* is the width of the RBC in the number of pixels. Examples of warping by S=8 are illustrated in Fig. 5(b).

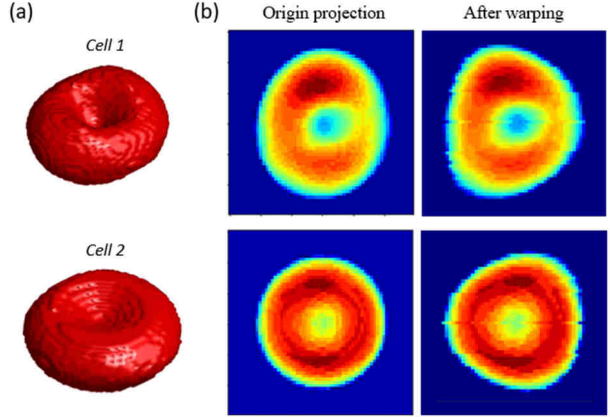

**Fig. 5** Two examples of (a) 3D rendered RI tomograms of RBCs, and (b) quantitative phase images obtained from direct projection of tomograms shown in (a) before (left) and after warping (right).

*2.7 RBC sample preparation and feature extraction*

A droplet of blood (about 2 µL) was obtained from one of the authors (Y. Chang) by finger prick and diluted in 0.85% phosphate-buffered saline (PBS). Part of the blood samples were diluted in PBS with 0.01% or 0.05% glutaraldehyde to generate stiffening effects and associated morphological changes in RBCs. Adding glutaraldehyde to PBS is known to cause an increase in osmolality, which in turn changes the morphology of RBCs. Therefore, the osmolality of the diluents containing glutaraldehyde was adjusted to be approximately the same as that of the PBS. After 20 minutes of treatment, the cells were extracted by centrifugation and diluted in PBS for imaging in the microfluidic channel.

Three types of features were obtained from reconstructed phase images or RI tomograms of RBCs, including morphology, intracellular content, and membrane stiffness. Morphological features directly quantified from 3D RI tomograms included each RBC's volume and surface area. To better characterize the shape of an RBC without being biased by its size, we also calculated sphericity by $\pi^{\frac{1}{3}}(6 \times volume)^{\frac{2}{3}}/(surface\ area)$. In addition, the mean diameter and biconcave disc parameters were quantified from 2D phase images of RBCs orientated at about θ=0°. Since RBCs do not contain membrane-bound organelles and consist mostly of hemoglobin molecules, they can be approximated as objects with homogeneous intracellular refractive index. The phase of a pixel (x, y) in an RBC phase image can be described as

$$\Delta\phi(x,y) = \frac{2\pi}{\lambda}[n_{RBC} - n_m]h(x,y), \tag{2}$$

where $h(x,y)$ is the local height of the RBC, $n_m$ is the refractive index of the diluent and $n_{RBC}$ is the average refractive index of the RBC. It follows that the phase of a pixel is proportional to the height at the local position under the homogeneity assumption. Therefore, 2D phase images of RBCs can be used to characterize the thickness profile or contour of the RBCs. We adopted the following equation [38],

$$\left(1 - \left(\frac{r}{R}\right)^2\right)^{\frac{1}{2}} \times \left(B_0 + B_2 \left(\frac{r}{R}\right)^2 + B_4 \left(\frac{r}{R}\right)^4\right), \tag{3}$$

to fit an RBC's phase expressed as a function of radial distance r, where *R* is the average radius, and $B_0$ is the phase at the RBC's center. Both parameters $B_2$ and $B_4$ decrease relatively to $B_0$ when the shape of an RBC flattens and deviates from a regular biconcave disc.

The intracellular content of RBCs is mostly hemoglobin. The refractive index of the cytosol of an RBC is linearly correlated with the mass density of biomolecules in it [39]. From a 2D phase image of an RBC, it is straightforward to calculate the total phase-area product of all pixels within the image by

$$\text{OV} = \frac{\lambda}{2\pi}\sum_{RBC}\Delta\phi(x,y)\,\Delta x \Delta y, \quad (4)$$

where OV is optical volume [40], and $\Delta x \Delta y$ is the area of each pixel in sample space. OV is a convenient parameter to assess the dry mass of cells from their 2D quantitative phase images. Precisely, suppose the contribution to intracellular RI by substances other than hemoglobin can be ignored in healthy RBCs. The dry mass of hemoglobin in an RBC can be estimated by OV/α where α is the specific refraction increment of hemoglobin [41]. Moreover, according to Eqs. (2) and (4), dividing OV by the physical volume of the identical RBC gives the difference between the average intracellular RI and the medium RI, $n_{RBC} - n_m$, which is proportional to the dry mass density of hemoglobin in the RBC [39].

The deformability or viscoelasticity of the RBC membrane was assessed by quantifying temporal fluctuations of the cell membrane using the same DHM instrument. After driving RBCs into the channel, the syringe pump was stopped for 10 minutes to allow RBCs to settle at the bottom of the channel. Then, we captured 300 interference images of RBCs at 180 frames/s. Quantitative phase images of the RBCs were reconstructed and cropped as described in Sec. 2.3. The temporal fluctuation in phase at each pixel was first quantified as the absolute difference between each image's phase value and the mean phase of all images, and an average fluctuation in phase was obtained over the whole duration of 300 images. That is, the mean membrane fluctuation in phase for each pixel (*x*, *y*) can be calculated as

$$\sigma_\phi(x,y) = \frac{\sum_t |\Delta\phi(x,y)_t - \Delta\phi(x,y)_{average}|}{T}, \quad (5)$$

where *t*=1,…*T* is the time index of each image, $\Delta\phi(x,y)_{average}$ is the average phase over the *T* images at the pixel. Finally, membrane fluctuations of all pixels in the image were averaged [42].

## 3 Results

### 3.1 Theoretical evaluation of two methods of RBC rotation angle determination

The proposed method for determining the orientation of RBCs was modified from the method proposed by Merola *et al.* [15]. We tested the two methods on 200 projections for each of the 15 RBCs to evaluate their performance in estimating the rolling angles. The mean errors in estimated rolling angles in the presence of total phase noise are 9.4° and 13.9° for the proposed method and previously reported method, respectively. The mean errors in the presence of total phase noise and image warping are 9.7° and 11.8° for the two methods, respectively. The proposed method showed significantly smaller errors (p<0.001) in the estimated rolling angles with or without the warping artifacts. The previously reported method uses normalized $C_4$ values over the whole sequence of phase images to determine the rolling angle, which is prone to errors since changes in intracellular mass distributions during the movement may alter the relation between $C_4$ and cosθ. The proposed method, on the other hand, divides the whole sequence of phase images into segments of 90° rolling. Normalizing $C_4+C_5$ values within each 90° segment is shown to more accurately recover the orientation of RBCs since image frames captured within each 90° segment are within 0.5 s and appear to have similar intracellular mass distributions. In addition, the inclusion of an additional $C_5$ term helps minimize the effects of nonideal shapes of RBCs during their movements on the fitting to the Zernike polynomials.

The performance of the proposed method was further evaluated by reconstructing 3D RI tomograms of the 15 RBCs from the 200 test phase images using the angles estimated. 3D morphological features, including the volume and surface of reconstructed tomograms, were used as target parameters. The gold standard was obtained by reconstructing 3D RI tomograms using 500 projected complex-field images with known and equally spaced rolling angles. Table 1 shows that the errors of the proposed method are smaller than those of the original method for estimating the rotation angle.

**Table 1** Comparison of errors in volume and surface area of reconstructed RBC tomograms between the proposed method and previously published method [15]

|  | Random rotation + Total phase variation | | Random rotation + Total phase variation + Image warping | |
|---|---|---|---|---|
|  | Volume | Surface area | Volume | Surface area |
| Our Method (N = 15) | 3.5 ± 3.3 % | 3.1 ± 1.9 % | 6.6 ± 3.3% | 3.9 ± 2.8% |
| $C_4$ estimated result (N = 15) | 10.8 ± 4.2 % | 5.4 ± 4.6 % | 10.29 ± 3.7% | 6.1 ± 3.3% |

(% error)

*3.2 Imaging healthy and glutaraldehyde-treated RBCs*

Reconstructed and unwrapped phase images of an RBC flowing through the FOV of the DHM setup are shown in visualization 1, demonstrating that the flow rate was appropriate to result in at least one complete revolution of RBCs within the FOV. It also shows that the speed of rolling was not constant. The maximum speed of RBCs' movement was found to be about 0.2 mm/s to maintain the natural shape of RBCs. Exemplary reconstruction results of RI tomograms of two healthy RBCs are shown in Fig. 6. The biconcave disc shape of the RBCs can be seen in both selected slices of the 3D tomograms and the surface-rendered graphs.

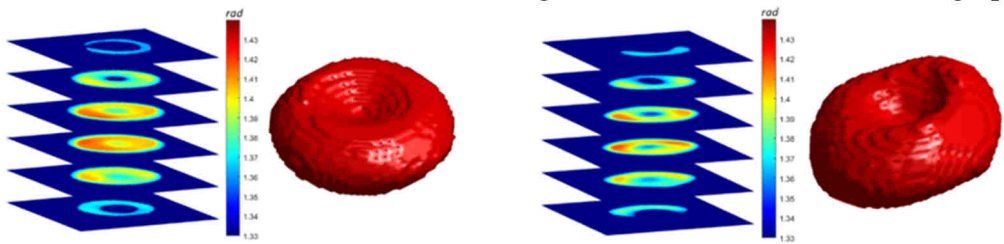

**Fig. 6** Two examples of reconstructed 3D RI tomograms of RBCs, showing slices at various depths on the left and 3D surface rendering on the right.

Influences of glutaraldehyde on RBCs were characterized by quantitative phase images and reconstructed RI tomograms of RBCs flowing through the microfluidic channel. Fig. 7(a)-(e) shows morphological features of RBCs in the three groups (control, 0.01% glutaraldehyde, and 0.05% glutaraldehyde). After the treatment, RBCs shrank in volume and mean diameter, as shown in Fig. 7(a) and 7(b), respectively. This is expected due to the crosslinking effects of glutaraldehyde on proteins [43] and has been measured by low-angle light scattering in flow cytometry [44]. Since both the surface area in Fig. 7(c) and the volume in Fig. 7(a) decreased after the treatment, it is uncertain how the shape of RBCs changed based on the measured surface area alone. Therefore, changes in RBC shape due to glutaraldehyde treatments were assessed with sphericity and biconcave disc parameters according to Eq. (3). As shown in Fig. 7(d) and 7(e) respectively, the sphericity increased and the sum of $B_2$ and $B_4$ decreased relatively to $B_0$ in treated cell populations, indicating that the RBCs' shape deviated from discocytes and resulted in a decreased concave depth. It is noted that concentrations of glutaraldehyde should be interpreted with caution. Fixation properties of glutaraldehyde in the same apparent concentration vary from batch to batch due to differences in the fraction of monomers and polymers [43]. Therefore, while similar trends in RBC characteristics due to glutaraldehyde treatments have been reported in the literature, the glutaraldehyde concentrations used may vary substantially.

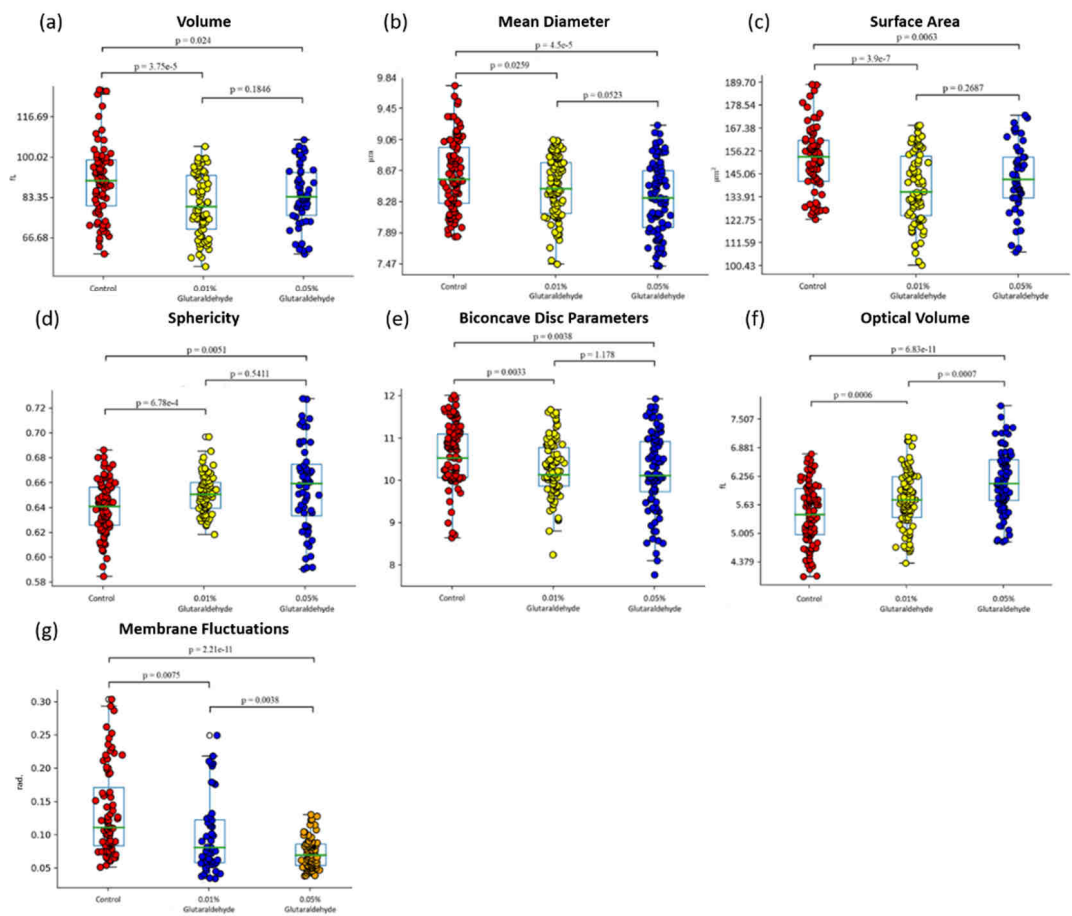

**Fig. 7** Comparison of features extracted from 3D RI tomograms and quantitative phase images of RBCs in the control group (sample size=108) and experimental groups treated with 0.01% (sample size=103) and 0.05% (sample size=95) glutaraldehyde, respectively. (a)-(e) Morphological features. (f) Optical volume or intracellular hemoglobin content. (g) Membrane fluctuation.

Fig. 7(f) shows that the OV of treated RBCs was higher than that of control RBCs, which is reasonable due to the addition of glutaraldehyde to RBCs that increases intracellular dry mass [43]. The decreased volume and increased intracellular dry mass in treated RBCs combined additively to increase the mass density in treated RBCs, which has been reported in [43]. In the control group results, the average OV of 5.4 fL corresponds to a mean hemoglobin dry mass of 26.6 pg, assuming a refraction increment of 0.203 ml/g for hemoglobin [45]. This value is within the range of typical mean corpuscular hemoglobin (MCH) for healthy human subjects. The ratio between the OV and the physical volume of each RBC in the control group results averaged about 0.06. This value is close to a previously reported average $n_{RBC} - n_m$ value of RBCs measured on samples from seven healthy human subjects [11]. The intracellular hemoglobin concentration could be calculated to be 30 g/dL by assuming the same refraction increment of 0.203 ml/g [45] and ignoring the contributions of other substances to the cellular dry mass. The estimated hemoglobin concentration was slightly smaller than typical values of mean corpuscular hemoglobin concentration (MCHC) in healthy subjects. The discrepancy may be attributed to individual variations in MCHC and refraction increments of hemoglobin, and differences in RBC volume measured by optical diffraction tomography and clinical blood analyzers [46].

The quantified membrane fluctuations are shown in Fig. 7(g) and indicate increased membrane stiffness in treated RBCs, which is expected since glutaraldehyde crosslinks proteins in the membrane and cytosol of RBCs [43].

## 4   Discussion

We implemented 3D imaging flow cytometry of RBCs based on optical diffraction tomography. It is label-free and quantifies 3D morphology and intracellular hemoglobin content of RBCs in their natural discocyte form, which are advantages over conventional blood smears. We demonstrated the capability of the developed cytometer to detect changes in the morphology, hemoglobin content, and deformability of RBCs due to glutaraldehyde treatments. The results indicate the potential of the implemented cytometer to detect changes commonly present in various RBC membrane disorders and metabolic syndrome. Recent evidence has also shown associations between decreased RBC deformability and cardiovascular disease conditions [24]. With further improvements for automated image reconstruction and analysis, the proposed cytometry could be a practical tool to provide fast measurements of important RBC parameters for the diagnosis, treatment monitoring, and management of various conditions.

The key to achieving high-throughput tomographic imaging of single RBCs is using self-rotation of continuously flowing cells to collect complex field images of the cells under multiple projection directions and quantifying the rotation angles from the acquired quantitative phase images [15]. This strategy has been adopted by [21] to image single and coagulated RBCs. We modified the previous method to improve the image quality. First, motion-induced noise known to plaque interferometry-based instruments is reduced by a common-path configuration for generating a uniform reference wavefront [25]. Compared to typical diffraction phase microscopy, where a pinhole is used as a spatial filter to generate the uniform reference beam [27], our method removes the pinhole and is more robust to misalignment. Second, the process for determining rotation angles of flowing RBCs by Zernike polynomial fitting [15] was modified to tolerate fluctuations in cellular shape and intracellular mass distributions during the movement of RBCs along the channel. Results show that the modified method improved accuracy in determining both rotation angles and the volume of reconstructed tomograms. Third, we considered effects of diffraction on the 3D reconstruction of RI tomograms using optical diffraction tomography [36], and applied total variation minimization to smooth intracellular RI distributions [47].

The throughput of the reported imaging flow cytometer is about 43 RBCs/min. It is currently limited by images of RBCs flowing through out-of-focus regions overlapping with those of in-focus RBCs due to incomplete confinement of RBCs to the depth of field by hydrodynamic focusing. This artifact could be reduced by manufacturing microfluidic channels with higher precision. In particular, decreasing the channel height to about twice the diameter of an RBC helps confine RBCs to the bottom half of the channel, and the laminar flow in the channel facilitates continuous rolling of the RBCs. The required precision could be achieved by photolithography with spin-coated photoresist for patterning the channels [20]. Although the flow speed cannot be significantly increased due to shape of RBCs, the throughput could be greatly improved by maximizing the number of flowing RBCs within the FOV of the DHM system. For example, one can increase the density of RBCs flowing in the channel and make the channel sufficiently wide so that the whole FOV is within the channel region.

## 5   Conclusion

To develop tomographic imaging flow cytometry for high-throughput single-cell characterization of RBCs, we designed and fabricated a microfluidic device to produce self-rotation of live RBCs flowing in the channel and complex field images of the RBCs were continuously acquired with a common-path off-axis digital holographic microscope. We modified the procedure of determining the rotation angle of RBCs for 3D reconstruction based on optical diffraction tomography. Results on simulated RBC projection images showed that the modified approach improved the accuracy in estimating the rotation angle and the volume of reconstructed RBCs as compared to the original method. The developed imaging flow cytometer was validated by imaging healthy RBCs, where the 3D morphology of RBCs was correctly reconstructed as biconcave discs. We further demonstrated the capability of the developed cytometer to quantify changes in RBC 3D morphology, hemoglobin content, and membrane fluctuations generated by glutaraldehyde treatments. The results of this study show the potential of the proposed cytometry to detect changes commonly seen in various RBC membrane disorders and metabolic syndrome.

*Disclosures*

All authors state that they have no relevant financial interests in this article and no other conflicts of interest to disclose.

*Acknowledgments*

The authors thank National Science and Technology Council in Taiwan for financial support (grant number NSTC 108-2221-E-002-081-MY3). The authors thank Prof. Nien-Tsu Huang of National Taiwan University for suggestions on the design and fabrication of the microfluidic device and Ms. Huai-Ching Hsieh for help with organizing and making figures and the video file.

Tomography Apparatus by Optimizing the Experimental Layout and Computational Processing, Cells 11 (2022) 2591.